\documentclass[12pt,twocolumn]{article}
\usepackage[a4paper,left=20mm,right=20mm,top=25mm,bottom=25mm,includeheadfoot]{geometry}

\usepackage{mathpazo}
\usepackage{graphicx}
\usepackage{amsmath,textcomp}
\usepackage{makeidx}
\makeindex

\usepackage{sectsty}
	\allsectionsfont{\sffamily\raggedright}
	\sectionfont{\sffamily\large\raggedright}
	\subsectionfont{\sffamily\normalsize\raggedright}
\usepackage{ftnright}

\usepackage{flushend}
\usepackage{cuted}
\usepackage{color}
\usepackage{graphicx}
\usepackage{hyperref}
\usepackage{pstricks,amssymb}

\usepackage{fancyhdr}
\begin{document}
%Do not change the \vspace command, which is needed to suppress extra space above the title.
\title{\vspace{-2em}\bfseries\sffamily What Hides within the Photograph:\\ the Analysis of a Light Curve in the Classroom}
\author{\normalsize H. Oll\'e${^1}$ and T. Kov\'acs${^2}$\\[2ex]
$^{1}$S\'ukromn\'e gymn\'azium s v.j.m, Dunajsk\'a Streda, Slovakia.\\
{\tt olle.hajnalka@mgds.eu}\\[2ex]
$^{2}$Institute of Physics, E\"otv\"os University,\\ P\'azm\'any P. s. 1A, 1117 Budapest, Hungary.\\
{\tt tamas.kovacs@ttk.elte.hu}
}

\date{\itshape Submitted on dd-MMM-yyyy}

\maketitle

\thispagestyle{fancy}

%Do not change the \sffamily command, it is needed to ensure that the abstract appears in a different font.
\begin{abstract}
{\sffamily
Data from the Kepler satellite were analysed using the Mikulski Archive for Space Telescopes (MAST) database. With the participation of 53 students, we determined the parameters of the HAT-P-7b (Kepler-2) exoplanet system (transit duration, planet-to-star radius ratio, orbital period, semi-major axis, and star mass). To achieve the result, we used approaches that are easy to understand and apply to secondary school students as well. In this way, they were able to gain insight into the essential process of light curve analysis.
}\\
\hrule
%Do not change the \hrule command, it is needed to separate the abstract form the main text.
\end{abstract}

\section{Introduction}
An everyday problem in education is that it is difficult to arouse the interest of the audience with traditional methods. We must not overlook this.In order to give a clear picture of the course of the research work, it is not necessary to present complicated procedures. A simple idea is a more expedient. The teacher must introduce methods and topics to the class that are able to satisfy the needs of the students, and at the same time make them think independently and recognize connections. Experience clearly shows that in addition to the results of current research, the method itself is interesting to them. Students are curious about the ways in which they can obtain information from a data set.

We tried to show them this in a database accessible to everyone. The website of the Mikulski Archive for Space Telescopes (MAST) \cite{kepler_db} provides an easy-to-use, publicly accessible interface for viewing and downloading, among other things, the light curves of the transiting exoplanet systems detected by the Kepler satellite and the associated flux data. Fortunately for us there are systems where the plane of the sky and the plane of the planet's orbit are at an angle close to 90$^{\circ}$ to each other, so when the planet passes in front of its central star, it obscures part of the disk, resulting in a decrease in light (dimmer) on the Kepler sensor plate (data). This kind of exoplanets are called transiting exoplanets. This reduction in light is (also) contained in said database, which can be easily used in education. By moving the cursor on the website \cite{kepler_db}, you can easily read the flux and detection time values that will be needed for further calculations. 

With our students, we tested how accurate the result could be by simply estimating the length and depth of the transit. Thus, avoiding the fitting of complex mathematical models, secondary school methods provide an illustrative picture of how astronomers determine the parameters of transiting planetary systems. The said estimations were made from a survey of 53 students which we will discuss in detail below. First we divided the students into smaller groups of 10-12 people determined by the capacity of the school IT classroom to provide each student with their own computer. Our primary goal was to determine the parameters of a selected exoplanet system - all 53 students observed the transit of the Kepler-2 system - as accurately as possible. It is important to mention that the students estimated the length and depth of the transit and the time between the two overlaps from the overlaps belonging to different time points. By averaging these, the final result was obtained, which was also used in further calculations.

 We should also mention that nowadays there are many options for bringing exoplanets into the classroom.
For instance: an example of how Williams \cite{williams} simulated the light curve of eclipsing binary stars in classroom. It is a simple, but great idea, that leads students to scientific thinking and attitudes. Different exoplanet detection methods such as radial velocity measurements can also be demonstrated in class as it has been done in \cite{lopresto2004,prather}. The Exoplanet Edu \cite{della-rose} package for iOS can generate theoretical light curves and radial velocity curves, and export worksheets for students. This gives a lot of opportunities in education and makes the lesson more interesting \cite{smartphone,george}. Another option is to use real light curves -- as is done in this study -- or radial velocity data of an available database to determine the parameters of the exoplanets. \cite{cowley,lopresto2005}

\section{The Kepler database}

Let us see how to get a desirable light curve to appear on the screen. By visiting the given link \cite{kepler_db}, do the following: 
\begin{itemize}
\item in the "Target Name" field, enter the ID of the desired object (for example, Kepler-2),
\item then click on the "Search" button to get a table,
\item here in the "Dataset Name" column you will find data series sorted by date. Clicking on virtually any of them will immediately bring up a light curve.
\end{itemize}

\begin{figure*}[!h]
\centering
\includegraphics[width=1.5\columnwidth]{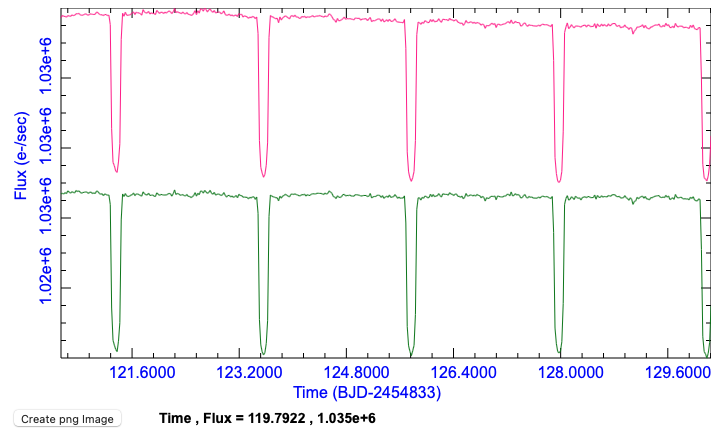}%[scale=0.4]{bs1}
%  \end{center}
\caption{\small Part of the Mikulski Archive for Space Telescopes database (screenshot)}
\label{fig:fig1}
\end{figure*}
The time is measured on the horizontal axis and incident flux on the vertical axis. What is immediately apparent that there is a green (SAP Flux - Simple Aperture Photometry Flux) and a pink curve (PDCSAP Flux - Pre-search Data Conditioning SAP Flux). Pink is a corrected flux that eliminates long-term trends based on knowledge and systematic errors \cite{kepler_db}. In our estimates, we examined this change in the already corrected flux. For a more accurate result, it is a good idea to zoom in on the figure. Simply select the desired area with the mouse and you can automatically zoom in on the detail. By moving the cursor, just below the figure, the time and flux values for that position are displayed. We need to record this data manually. For simplicity, we have created an Excel spreadsheet \cite{mgds} that is easy for students to use. Entering each data here will automatically calculate the system parameters.

\section{Determining the physical parameters}

By studying the light curves of the planetary systems, we can obtain information about the radius of the planet and the star, the orbital inclination of the planet's orbital plane (impact parameter), the average density of the star, the orbital period, and the semi-major axis of the orbit \cite{seager2003}. The above-mentioned properties of the exoplanet system may carry information indicating the presence of additional planets or moons, such as changes in orbit elements or anomalies in the form of the light curve \cite{seager2003}. It is clear that some effects appear on the light curve to such a small extent that they cannot be detected by such simple methods.
For us, the transit duration, the radius of the star and its planet, the orbital period, the semi-major axis of the orbit, the mass of the star are important. Let us look at them in turn!

\section{Transit duration:}

As an example, consider the light curve of the planet HAT-P-7b (the Kepler ID is: Kepler-2). Referring to Fig. \ref{fig:fig1}, the time of the beginning of the transit ($\tau_1$) and the end ($\tau_2$) can be immediately estimated. Subtracting these two data gives the length of the transit ($\tau$) measured in days. In our case, this value becomes 0.1667 $\pm$ 0.009 days, while the literature value is 0.1669 $\pm$ 0.003 days \cite{pal2008}.

\begin{figure}[!h]
\centering
\includegraphics[width=1.0\columnwidth]{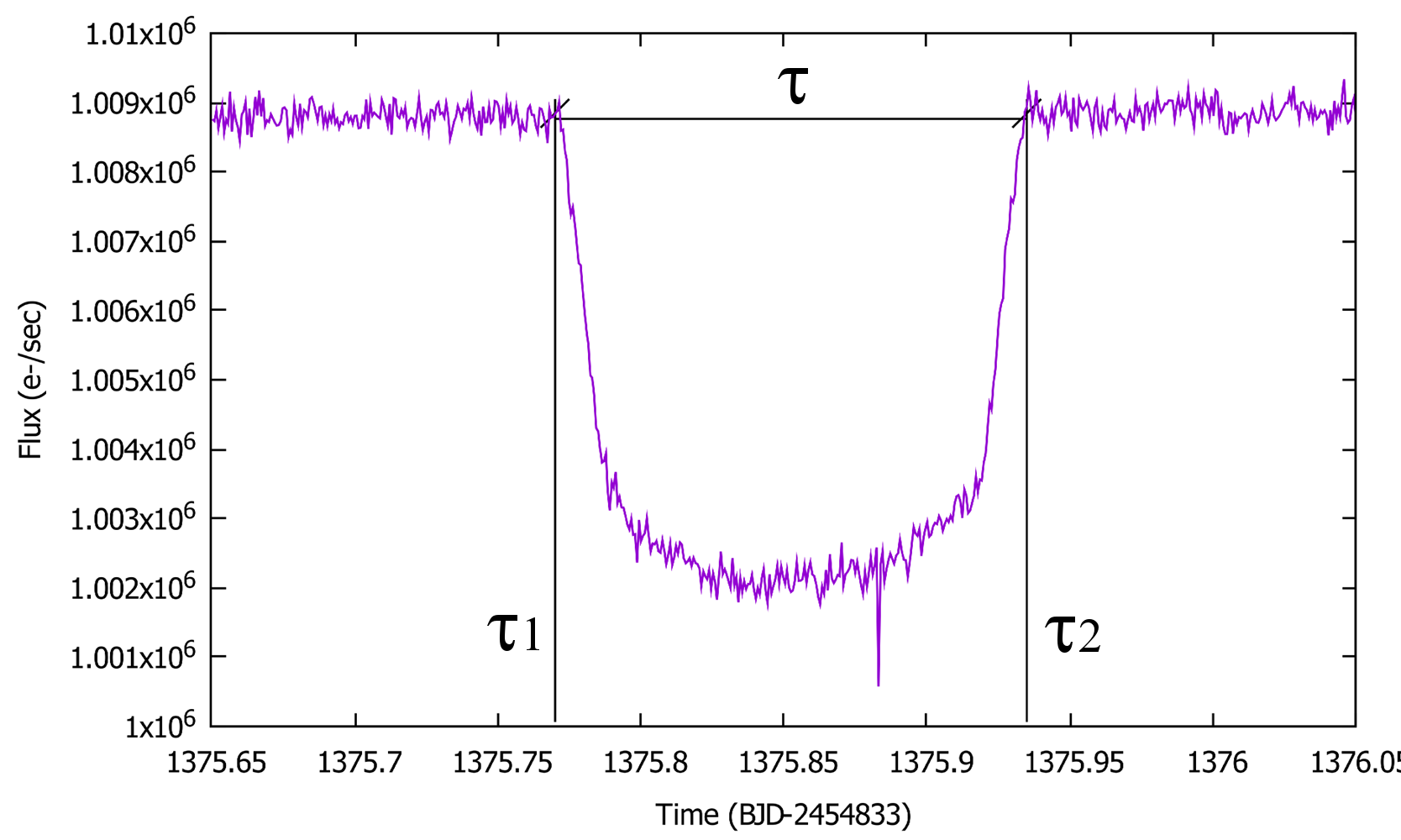}%[scale=0.4]{bs1}
%  \end{center}
\caption{\small The decrease in light caused by planet HAT-P-7 b, from the site of the MAST database mentioned in the introduction}
\label{fig:fig2}
\end{figure}

\section{Planet-to-star radius ratio:}
When the planet obscures part of the star, the observed brightness decreases. From a geometric point of view, we get that the ratio of the apparent disk of a star to the visible disk of a transiting planet gives the relative luminous loss ($\Delta F$). From this measure, the radius ratio of the planet to the central star can be easily determined \cite{seager2003}:

\begin{equation}
    \Delta F=\frac{F_{\mathrm{max}}-F_{\mathrm{min}}}{F_{\mathrm{max}}}=\frac{\pi r^2}{\pi R^2},
\end{equation}
where $r$ is the radius of the planet and $R$ is the radius of the central star. So
\begin{equation}
\frac{r}{R}=\sqrt{\Delta F}.
\end{equation}

From the flux reduction, students determined the radius ratio of the system, for which they received 0.0807 $\pm$ 0.0036. The literature value of the HAT-P-7b radius ratio, i.e. 0.077590 $\pm 3\times 10^{-5}$ \cite{pal2008}.

In this case, we also examined the system using another method. It is also possible to download the measurement data from the MAST website. To make this data easy to use for students, we have created a short Python program and a description that makes it easier to understand \cite{mgds}. The program creates an editable Excel file output, which we can also analyze.
It will certainly be a challenge for students to work with a data set of this size. It is an essential part of education to teach students to work with such large data sets \cite{mgds}. It forces them to invent methods and procedures for easier handling. By looking through the data systematically and finding the transits, we can determine the maximum flux between the two transits and the minimum flux during the transit by a simple average calculation. In this way, a radius ratio of 0.0772 $\pm$ 0.0052 was obtained, which is a fairly good result with secondary school apparatus.

\section{Orbital period and view of the trajectory:}

To determine the orbital period, two consecutive transits must be detected, since then the planet is in the same position during successive revolutions. Fortunately, we can easily access this on the website we are using.

\begin{figure}[!h]
\centering
\includegraphics[width=1.0\columnwidth]{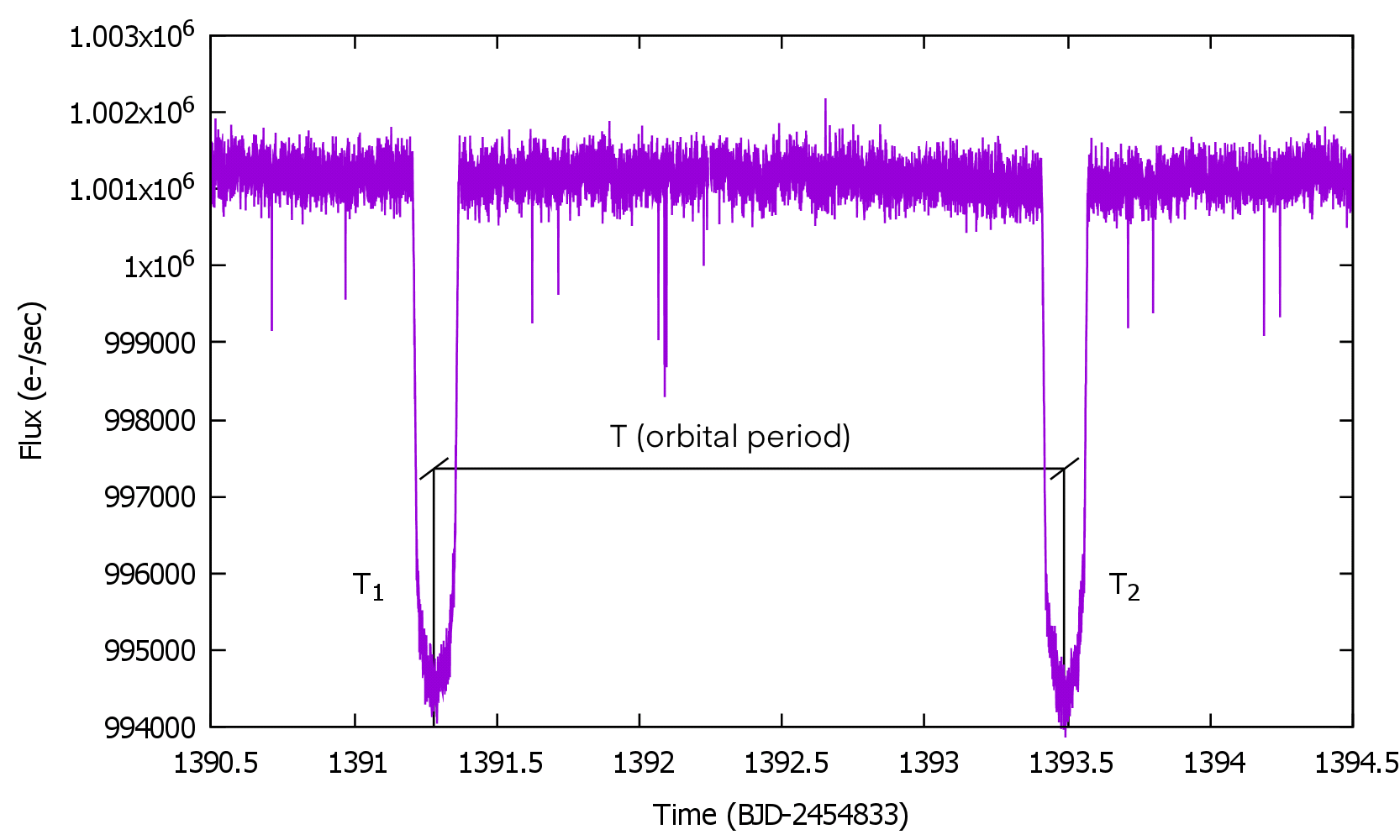}%[scale=0.4]{bs1}
%  \end{center}
\caption{\small To determine the orbital period, we need two consecutive transits at the same time.}
\label{fig:fig2}
\end{figure}

To determine the orbital period, adjust the view in a way that two consecutive transits shall appear, then move the cursor to read the time T between the two minima $T_1$ and $T_2$, which we find 2.201631 $\pm$ 0.011335 days (Fig. \ref{fig:fig2}). This is again a good estimate, as the literature value is $2.204737 \pm 1.7 \cdot 10^{-5}$ days \cite{pal2008}.

We now determine the orbit of the same system, i.e., the average distance between the planet and its star. Because the eccentricity of the Kepler-2 (HAT-P-7b) is very small \cite{pal2008}, it was a minor deviation to assume the planetary orbit to be circular.

\begin{figure}[!h]
\centering
\includegraphics[width=1.0\columnwidth]{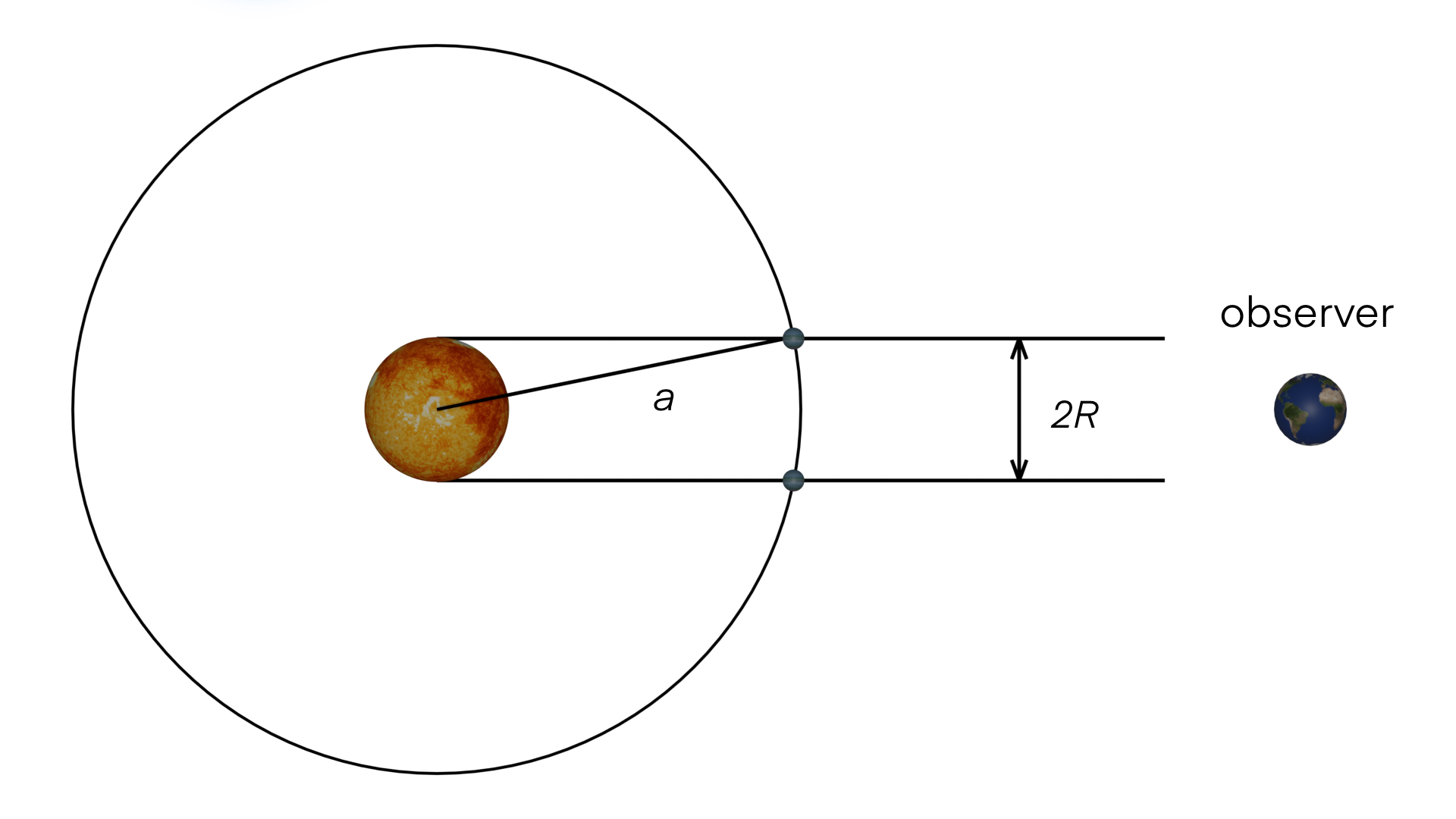}%[scale=0.4]{bs1}
%  \end{center}
\caption{\small A face-on view of the planet's orbital plane}
\label{fig:fig3}
\end{figure}

With simple secondary school level physics, it can be seen that the orbital velocity is written as the quotient of the circumference of the circle and the period:  
\begin{equation}
    v=\frac{2\pi a}{T},
\end{equation}
where $T$ is the aforementioned orbital period,  $a$ is the orbital radius (semi-major axis), and $v$ is the orbital velocity. It can also be seen from Fig. \ref{fig:fig3} that the orbital velocity can be approximated by a relation that contains only the radius R of the star and the length  ($\tau$) of the transit: 

\begin{equation}
    v=\frac{2R}{\tau}.
\end{equation}

If we match the two equations and express the radius of the orbit, we get the following relation: 
\begin{equation}
    a=\frac{RT}{\pi\tau}.
\end{equation}
The problem is that only the radius ratios can be determined from the light curve, so we still need the specific radius of the star. The required data for some planetary systems are given in Table 2. However, entrusting information retrieval of these data to students can also be an excellent task, which can be instructive in many ways. They could be struck by the fact that literary values are also constantly updated, they learn to choose from the available sources, and they get much more information while searching.

According to the students' estimate the semi-major axis is 0.0366$\pm$0.0011 AU, while the currently accepted value is 0.03796 $\pm$ 0.00063 AU \cite{pal2008,hatp7}.

\section{The mass of the star:}
Knowing the radius of the orbit (semi-major axis), we estimate the mass of the star according to Kepler's third law \cite{umass}. It should be noted that, for the sake of simplicity, we consider the mass of the planet to be negligibly small, it does not appear in the equation. The stellar mass can be determined by the following relationship: 
\begin{equation}
    M_{\star}=\frac{4\pi^2a^3}{GT^2},
\end{equation}
where $M_{\star}$ is the mass of the star in kilograms, $a$  is the semi-major axis of the orbit in meters, $G$ is the gravitational constant (its value is $6.6743\times 10^{-11}m^3kg^{-1}s^{-2}$), and $T$ is the orbital period in seconds. The values obtained from the light curve were replaced in the formula. The mass of the star was thus 1.4823 $\pm$ 0.4107 solar mass, in reality it is 1.500 $\pm$ 0.03 solar mass \cite{pal2008,hatp7}.

\section{Summary}

While solving the problem, be sure to make students aware that these methods provide only an illustrative picture of how we can extract information from light curve analysis, but the results obtained are only approximate. Therefore, in addition, it should be compared with the actual data in the databases.

From a didactic point of view, it is important because the problem can be approached from several perspectives, thus giving a complex picture of the system. Celestial mechanical factors, geometric arrangement, photometric properties must be taken into account. In fact, we only had to find a single piece of data, the true radius of the star, everything else we determined from the light curve itself.
The analysis of light curves could, of course, be continued. It should be taken into account that in multi-planet systems, the planets also perturb each other's orbits, resulting in the orbital elements changing in secular time scales. Here a clear relationship can be found between the change in transit length (or the change in time between two transits) and the change in orbital elements.
The parameters of the HAT-P-7b (Kepler-2) system were determined with the involvement of fifty-three students. Compared to the literature \cite{pal2008}, we obtained very good results for transit duration, planet-to-star radius ratio, orbital period, semi-major axis of the orbit, and star mass. It is clear that the method is suitable for determining certain parameters of planetary systems, which is also summarized in Table 1.

To facilitate further work, we have collected some planetary systems (Table 2) that are suitable for use in the presented method. During the compilation, it was taken into account that the relative flux reduction of the coverage is large and the orbital period is relatively small, so the transits should be frequent.

Astronomer Oll\'e Hajnalka is a PhD student at the E\"otv\"os Lor\'and University Doctoral School of Physics and a physics teacher at the Private Secondary School in Dunaszerdahely. She considers it important for her students to become acquainted with modern research methods and to show the practical use of their acquired knowledge.

Tam\'as Kov\'acs is an astronomer, an employee of the Institute of Physics at E\"otv\"os Lor\'and University. He wrote his doctoral dissertation on celestial mechanics. His current areas of interest are nonlinear dynamical systems, time series analysis, complex networks, and the statistical physical description of phase-space transport and their astronomical applications. Bolyai scholarship holder. He is an active supervisor of the Doctoral School of Physics.

\begin{table*}
%\begin{widetext}
\begin{center}
\begin{tabular}{|c|c|c|}
\hline 
Parameter & Values estimated by the students & Values within the literature \\
\hline 
$\tau$ & 0.1667$\pm$0.009 days &  0.1669$\pm$0.003 days\\
\hline 
$r/R$ & 0.0807 $\pm$ 0.0036 (from the figure)
& 0.077590 $\pm$ 3$\times$10-5\\
\hline 
& 0.0772 $\pm$ 0.0052 (from the data) & \\
\hline
$T$ & 2.201631$\pm$0.011335 days &2.204737 $\pm$ 1.7$\times$10$^{-5}$ days\\
\hline
$a$ & 0.0366$\pm$0.0011 AU & 0.03796 $\pm$ 0.00063 AU\\
\hline
$M_{\star}$ & 1.4823$\pm$0.4107$M_{\odot}$ &1.500 $\pm$ 0.03 $M_{\odot}$\\
\hline
\end{tabular}  \\
\caption{\small Data estimated by the students for the HAT-P-7b (Kepler-2) system. Comparing them with the values in the literature [8]
it is clearly visible that the method is applicable for determination of certain system parameters.}
\end{center}
%\end{widetext}
\end{table*}

\begin{table}
\begin{center}
\begin{tabular}{|c|c|}
\hline 
Name & $R_{\star}$($M_{\odot}$)\\
\hline 
Kepler-1 & 1.003 $\pm$ 0.033\\
\hline
Kepler-7 & 1.843+0.048-0.066\\
\hline
Kepler-12 & 1.483+0.025-0.029\\
\hline
Kepler-15 & 0.9920.058-0.07\\
\hline
Kepler-41 & 0.966 $\pm$ 0.032\\
\hline
Kepler-43 & 1.420 $\pm$ 0.07\\
\hline
Kepler-45 & 0.55 $\pm$ 0.11\\
\hline
\end{tabular}  \\
\caption{\small Systems suitable for the testing of our method, which have a relatively large decrease in flux. We also provided the value necessary for the calculation of the semimajor-axis, the radius of the central star [8].}
\end{center}
\end{table}
\section*{Acknowledgments}
We thank the anonymous referee for the helpful comments, and constructive remarks on this manuscript.

\end{document}